\documentclass[pra,aps,10pt,a4paper]{revtex4-2}

\usepackage{graphicx}
\usepackage{dcolumn}
\usepackage{bm}
\usepackage{CJKutf8}
\usepackage{xcolor}

\begin{document}
\begin{CJK*}{UTF8}{gbsn}
\title{The optimal strategy of two-photon interferometric sensing in diverse noise environments}


\author{Teng-fei Yan$^{1}$}

\author{Zhuo-zhuo Wang$^{1}$}

\author{Qi-qi Li$^{1}$}

\author{Peng-long Wang$^{1}$}

\author{Rui-Bo Jin$^{2,3}$}

\author{Bai-hong Li$^{1}$}
\email{li-baihong@163.com}




\affiliation{$^{1}$
School of Physics and Information Science, Shaanxi University of Science and Technology, Xi’an 710021, China
}%

\affiliation{$^{2}$Key Laboratory of Low-Dimensional Quantum Structures and Quantum Control of Ministry of Education, Department of Physics and Synergetic Innovation Center for Quantum Effects and Applications, Hunan Normal University, Changsha 410081, China}
\affiliation{$^{3}$Hubei Key Laboratory of Optical Information and  Pattern Recognition,  Wuhan Institute of Technology, Wuhan 430205, China\\}

\date{\today}

\begin{abstract}
Quantum sensing based on two-photon interferometry manifests quantum superiority beyond the classical precision limit. However, this superiority is usually diminished inevitably by the noise. Here, we analyze the sensitivity of two typical two-photon interferometries to the noise, that is, Hong-Ou-Mandel (HOM) and N00N state interferometry. It is found that HOM (N00N state) interference, which depends on the biphoton frequency difference (sum), is insensitive (sensitive) to the phase noise in both the manners of spectrally non-resolved and resolved detections in practice, suggesting their potential applications of sensing for different noise scenarios. Furthermore, spectrally resolved detection outperforms spectrally non-resolved one for the two interferometries, especially for the scope that exceeds the coherence time of biphotons. The findings provide an optimal strategy for the practical applications of two-photon interferometric sensing in diverse noise environments.



\end{abstract}


\maketitle


\section{\label{sec:1}INTRODUCTION}
Quantum sensing, which utilizes the nonclassical quantum resources, such as the entangled or squeezed states, has shown quantum superiority in sensing precision and resolution beyond the classical limit \cite{RMD-2017,NP-2018}. Due to these remarkable capabilities, it has found widespread applications in high-precision magnetometery \cite{PRAPP-2020,AQT-2024,PRX Quantum-2024}, biological microscopy \cite{NP-2013,nature-2021,NRP-2023}, gravitational-wave detection \cite{LRR-2019,PRD-2023,NC-2024} and so on. In the field of photonic quantum \cite{NP-2018,JLT-2015}, quantum sensing can be realized by some typical quantum interferometers, such as Hong-Ou-Mandel (HOM) and N00N state interferometers \cite{Jin2024review}.

HOM interference \cite{PRL1987,RPP2021,Jin2024review}, a well-known non-classical interference effect, has many important applications in advanced quantum technologies, and in particular it has been shown to have powerful metrological potential for high-precision imaging and sensing applications \cite{NATURE2001,PRL2006,NP2011}. As a result, the HOM interferometer has become a valuable apparatus for quantum parameter estimation \cite{PRL2007,SA2018,npj2019,Cyril2023,PRA2020,PRA2021,Fabre2021PRA,PRA2022,CHENPRAPP2023,PRA2023,
PRL2023,PRL2024,CHENPRAPP2023,PRAPP2023,SA2025,OE-2025,Chen-OE-2025}. The N00N state interferometer, enabling phase measurements at the Heisenberg limit, has been widely used in quantum lithography \cite{NOON,NOON1}, quantum high-precision phase measurement \cite{NOON2,NOON3,NOON4}, quantum microscopy \cite{microscopy1,microscopy2,microscopy3}, error correction \cite{error}, and so on. 


In general, the performance of quantum sensing based on quantum interferometry is diminished inevitably more or less by the environmental noise \cite{PRA2021,NP-2011,PRL-2015,PRL-2017}. In practice, different quantum interferometry may have diverse noise sensitivity, resulting in their different application scenarios of quantum sensing under different noise environments.



In this paper, we explore the sensitivity of two typical two-photon interferometries to the noise, that is, HOM and N00N state interferometry, by analyzing Fisher information (FI) obtained from their coincidence probabilities.  It is found that HOM (N00N state) interferometry, which depends on biphoton frequency difference (sum), is insensitive (sensitive) to the phase noise in both manners of spectrally non-resolved and resolved detections in practice. This suggests that the two interferometries have different application scenarios of quantum sensing; that is, HOM interferometry can be used in a high-noise scenario, while N00N state interferometry can only be used in a low-noise one. Furthermore, spectrally resolved detection outperforms spectrally non-resolved one for the two interferometries, especially for the scope beyond the coherence time of biphotons. The findings provide an optimal strategy for the practical applications of two-photon interferometric sensing in diverse noise environments.

The remainder of the paper is organized as follows. In Sec. \ref{sec:2}, we investigate the sensitivity of quantum sensing based on spectrally non-resolved (Subsection A) and resolved (Subsection B) HOM interferometry on the phase noise by analyzing FI obtained from their coincidence probabilities. In Sec. \ref{sec:3}, we investigate the sensitivity of the phase noise to quantum sensing based on spectrally non-resolved (Subsection A) and resolved (Subsection B) N00N state interferometry. Both sections derive the ultimate quantum sensing limits achievable under the ideal conditions of zero loss and perfect visibility and also present a comparative analysis of the results obtained from spectrally non-resolved and resolved detections. Finally, Section \ref{sec:conclude} summarizes the findings and concludes the paper.

\section{\label{sec:2} Quantum sensing based on two-photon Hong-Ou-Mandel interferometry}
\subsection{\label{A} Quantum sensing based on spectrally non-resolved Hong-Ou-Mandel interferometry}

According to the principle of quantum metrology \cite{PRL2007}, there exists a fundamental limit to the precision of the estimation of a parameter, e.g., the time delay, which is determined only by the probe state used, regardless of the details of the final measurement step \cite{SA2018,npj2019,Fabre2021PRA,CHENPRAPP2023,OE-2025}. This is the so-called Quantum Cramér-Rao bound (QCRB) \cite{Helstrom}. Two-photon interferometry is a specific measurement strategy to saturate the QCRB. Through the FI analysis, one can identify whether such a strategy can saturate the QCRB or not.  For quantum sensing, such a bound also determines the limit of the sensing precision and resolution, which can be obtained by the FI analysis as well.

First, let us consider the sensitivity of the phase noise to quantum sensing based on spectrally non-resolved HOM interferometry under a noise environment. Adopting the noise model from Ref.\cite{PRA2021}, there are three possible measurement outcomes for a real HOM interferometer: either both photons are detected, one photon is detected, or no photon is detected, which correspond to three probability distributions, \cite{SA2018},
\begin{eqnarray}
\label{R123}
&&R_2(\tau)=\frac{1}{2}(1-\gamma)^2\Big(1-Ve^{-2\sigma_{-}^2(\tau-\epsilon)^2}\Big),
\nonumber\\
&&R_1(\tau)=\frac{1}{2}(1-\gamma)^2\Big(\frac{1+3\gamma}{1-\gamma}+Ve^{-2\sigma_{-}^2(\tau-\epsilon)^2}\Big),\nonumber\\
&&R_0(\tau)=\gamma^2.
\end{eqnarray}
where subscripts 0, 1, and 2 denote the number of detectors that register a click, corresponding to total loss, bunching, and coincidence, respectively. $\gamma$ and $V$ represent photon loss and imperfect experimental visibility \cite{SA2018,npj2019,OE-2025}, respectively. In Eq.(\ref{R123}), the probability distributions are expressed as a function of the time delay $\tau$ between two arms of the interferometer, and $\sigma_{-}$ denotes the RMS (root mean square) bandwidth of a Gaussian entangled state used, which is determined by the phase-matching function associated with biphoton frequency difference \cite{JIN-Optica-2018,Li-PRA-2023,OLT-2025,OE-2025}.

It is found that Eq.(\ref{R123}) introduces a time shift caused by the frequency-dependent phase shift $\epsilon$, and a constant phase $\theta$ does not affect the probability distribution. Note that the coincidence probabilities in Eq.(\ref{R123}) are expressed in a manner of spectrally non-resolved detection, since it comes from the frequency integral of the original formula \cite{JIN-Optica-2018,Li-PRA-2023,OLT-2025}. If the detectors have high enough resolution to resolve the frequency of each photon so that the frequency integral can be removed, this is called spectrally resolved detection, as will be seen in Part B.


According to the noise model in Ref.\cite{PRA2021},
\begin{eqnarray}
\label{J}
J_{\epsilon}(\varepsilon)=\frac{e^{-\varepsilon^2/2\eta_{\epsilon}^2}}{\sqrt{2\pi} \eta_{\epsilon}},
J_{\theta}(\vartheta)=\frac{e^{-\vartheta^2/2\eta_{\theta}^2}}{\sqrt{2\pi} \eta_{\theta}}.
\end{eqnarray}
where $\epsilon$ represents a frequency-dependent phase shift and $\theta$ a frequency-independent one. Such phase noises might arise from vibrations or heating in the system. Correspondingly, $\eta_{\epsilon}$ and $\eta_{\theta}$ control the width of the Gaussian weighting factors, the strength of these noise processes.

The probabilities in Eq.(\ref{R123}) after considering noise distributions above can be rewritten by
\begin{eqnarray}
\label{RJ}
R_{i}^{\eta}=\int_{-\infty}^{\infty}\int_{-\infty}^{\infty}d\vartheta d\varepsilon J_{\epsilon}(\varepsilon)J_{\theta}(\vartheta)R_i.
\end{eqnarray}
Substituting Eqs.(\ref{R123}) and (\ref{J}) into Eq.(\ref{RJ}), we have
\begin{eqnarray}
\label{R123-noise}
&&R_2^{\eta}(\tau)=\frac{1}{2}(1-\gamma)^2\Big(1-(V/\sqrt{A})e^{-2\sigma_{-}^2\tau^2/A}\Big),
\nonumber\\
&&R_1^{\eta}(\tau)=\frac{1}{2}(1-\gamma)^2\Big(\frac{1+3\gamma}{1-\gamma}+(V/\sqrt{A})e^{-2\sigma_{-}^2\tau^2/A}\Big),\nonumber\\
&&R_0^{\eta}(\tau)=\gamma^2.
\end{eqnarray}
where $A=1+4\sigma_{-}^2\eta_{\epsilon}^2.$ If there is no noise, that is, $\eta_{\epsilon}=0$, then $A=1$, and Eq.(\ref{R123-noise}) will reduce to the original formula \cite{JIN-Optica-2018,OLT-2025}. It is found that HOM interferometry is influenced only by frequency-dependent phase shift $\epsilon$ and is independent of $\theta$. The introduction of the phase noise $\eta_{\epsilon}$ degrades the visibility of HOM interference by a factor of $\sqrt{A}$.

\textcolor{red}{} The outcome probabilities $R_i^{\eta}(\tau)$ in Eq.(\ref{R123-noise}) can now be used to construct an estimation of the sensing parameter of the time delay $\tau$. According to classical estimation theory, there exists a fundamental limit to determine such a sensing parameter, which is lower-bounded by
\begin{equation}
\label{CR}
\delta \tau_{CRB}=\frac{1}{\sqrt{NF_{\tau}}} \geq \delta \tau_{QCRB},
\end{equation}
where $N$ is the number of repetitions of the experiment, and $F_{\tau}$ is the FI. Under the noise environment, it refers to $F_{\tau}^{\eta}$, which reads
\begin{equation}
\label{FI}
F_{\tau}^{\eta}=\sum_{i}\frac{(\partial_{\tau} R_i^{\eta}(\tau))^2}{R_i^{\eta}(\tau)}=\frac{(\partial_{\tau}R_2^{\eta}(\tau))^2}{R_2^{\eta}(\tau)}+\frac{(\partial_{\tau}R_1^{\eta}(\tau))^2}{R_1^{\eta}(\tau)}+\frac{(\partial_{\tau}R_0^{\eta}(\tau))^2}{R_0^{\eta}(\tau)}.
\end{equation}
Substituting Eq.(\ref{R123-noise}) into Eq.(\ref{FI}), we finally arrive at
\begin{eqnarray}
\label{FI-HOM}
F_{\tau}^{\eta}=\frac{8V^2 \tau^2 \sigma_{-}^4(1-\gamma)^{2} e^{-4\sigma_{-}^2\tau^2/A}}{A^3} \Big(\Big(1-Ve^{-2\sigma_{-}^2\tau^2/A}/\sqrt{A}\Big)^{-1}+\Big(\frac{1+3\gamma}{1-\gamma}+Ve^{-2\sigma_{-}^2\tau^2/A}/\sqrt{A}\Big)^{-1}\Big),
\end{eqnarray}
Eq.(\ref{FI-HOM}) can be used to analyze the sensitivity of the FI on the phase noise $\eta_{\epsilon}$ by a factor of $A$ for spectrally non-resolved HOM interferometry.  Under the ideal conditions of zero loss and perfect visibility, i.e., $\gamma=0,V=1$, Eq.(\ref{FI-HOM}) can be simplified to
\begin{eqnarray}
\label{FI-ideal}
F_{\tau,ideal}^{\eta}=\frac{16\sigma_{-}^4\tau^2}{A^2(Ae^{4\sigma_{-}^2\tau^2/A}-1)},
\end{eqnarray}
If there is no noise, i.e., $\eta_{\epsilon}=0$, then
\begin{eqnarray}
\label{FI-ideal-1}
F_{\tau,ideal}^{\eta}=\frac{16\sigma_{-}^4\tau^2}{e^{4\sigma_{-}^2\tau^2}-1},
\end{eqnarray}
This limit is known as Cramér-Rao bound (CRB) \cite{Cramer}, which is associated with a particular quantum state and a specific measurement strategy. The QCRB can be obtained by maximizing over all possible measurements on the probe state. By evaluating the FI for this set of probabilities, we find that its upper bound is achieved under the ideal case at the zero delay point, given by
\begin{equation}
\label{FI0}
\lim_{\tau \to 0}F_{\tau,ideal}^{\eta}=4\sigma_{-}^2.
\end{equation}
The result is consistent with those obtained in \cite{SA2018,npj2019,Fabre2021PRA,CHENPRAPP2023,OE-2025,SA2025}. It is clear that we recover the QCRB under the ideal case at the zero delay point, thus obtaining the sensing limit of the time delay $\tau$ based on spectrally non-resolved HOM interferometry.

\subsection{\label{B} Quantum sensing based on spectrally resolved Hong-Ou-Mandel interferometry}
Now, let us consider the sensitivity of the phase noise to quantum sensing based on spectrally resolved HOM interferometry under a noisy environment. If the spectral resolution of the detectors is high enough so that they can resolve the frequency of each photon. In this case, the coincidence probability can be obtained directly by removing the frequency integral in the original formula \cite{JIN-Optica-2018,Li-PRA-2023,OLT-2025}, which can be written as
\begin{eqnarray}
\label{R123-SR}
&&R_2^{\eta}(\tau,\omega_-)=\frac{(1-\gamma)^2}{2(2 \pi \sigma_{-}^2)^{\frac{1}{2}}}e^{-\omega_{-}^2/8\sigma_{-}^2}\Big(1-V\cos[\omega_-(\tau-\epsilon)]\Big),
\nonumber\\
&&R_1^{\eta}(\tau,\omega_-)=\frac{(1-\gamma)^2}{2(2 \pi \sigma_{-}^2)^{\frac{1}{2}}}e^{-\omega_{-}^2/8\sigma_{-}^2}\Big(\frac{1+3\gamma}{1-\gamma}+V\cos[\omega_-(\tau-\epsilon)]\Big),\nonumber\\
&&R_0^{\eta}(\tau,\omega_-)=\frac{\gamma^2}{(2 \pi \sigma_{-}^2)^{\frac{1}{2}}}e^{-\omega_{-}^2/8\sigma_{-}^2}.
\end{eqnarray}
where $\omega_-=\omega_s-\omega_i$ represents the biphoton frequency difference between the signal and the idler photons.

\begin{figure}[th]
\begin{picture}(380,160)
\put(0,0){\makebox(375,150){
\scalebox{0.54}[0.54]{
\includegraphics{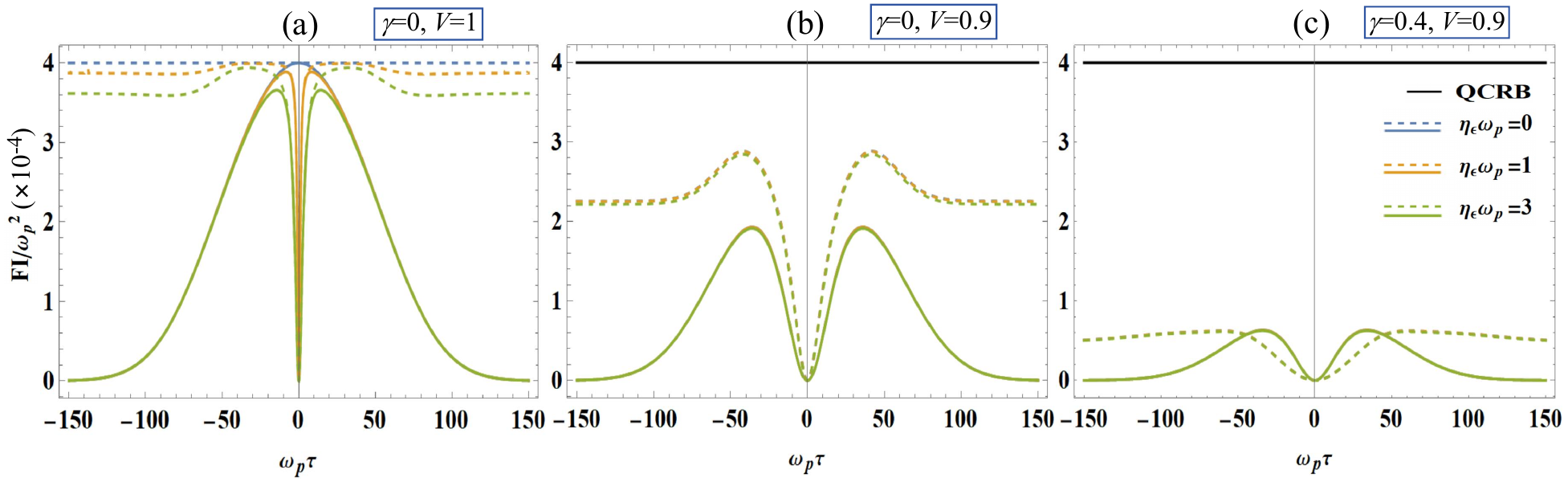}
}}}
\end{picture}
\caption{\label{Fig1}
Fisher information, in units of $\omega_{p}^2$, as a function of $\omega_{p} \tau$ for (a) $\gamma=0,V=1$, (b) $\gamma=0,V=0.9$, and (c) $\gamma=0.4,V=0.9$. The dashed and solid lines represent spectrally resolved and non-resolved results, respectively, when $\eta_{\epsilon}\omega_p$=0 (blue), 1 (orange), and 3 (green). The bold black lines represent the QCRB.}
\end{figure}

After using the noise model in Eqs.(\ref{J}), and (\ref{RJ}), we have
\begin{eqnarray}
\label{R123-SR}
&&R_2^{\eta}(\tau,\omega_{-})=\frac{(1-\gamma)^2}{2(2 \pi \sigma_{-}^2)^{\frac{1}{2}}}e^{-A\omega_{-}^2/8\sigma_{-}^2}\Big(e^{\omega_{-}^2\eta_{\epsilon}^2/2}-V\cos(\omega_-\tau)\Big),
\nonumber\\
&&R_1^{\eta}(\tau,\omega_{-})=\frac{(1-\gamma)^2}{2(2 \pi \sigma_{-}^2)^{\frac{1}{2}}}e^{-A\omega_{-}^2/8\sigma_{-}^2}\Big(e^{\omega_{-}^2\eta_{\epsilon}^2/2}\frac{1+3\gamma}{1-\gamma}+V\cos(\omega_{-}\tau)\Big),\nonumber\\
&&R_0^{\eta}(\tau,\omega_{-})=\frac{\gamma^2}{(2 \pi \sigma_{-}^2)^{\frac{1}{2}}}e^{-A\omega_{-}^2/8\sigma_{-}^2}.
\end{eqnarray}

The corresponding FI can be calculated as \cite{CHENPRAPP2023,PRAPP2023}
\begin{equation}
\label{FI-SR}
F_{\tau}^{\eta}(\tau)=\int \Big(\frac{(\partial_{\tau}R_{2}^{\eta}(\tau,\omega_{-}))^2}{R_{2}^{\eta}(\tau,\omega_{-})}+\frac{(\partial_{\tau}R_{1}^{\eta}(\tau,\omega_{-}))^2}{R_{1}^{\eta}(\tau,\omega_{-})}+\frac{(\partial_{\tau}R_{0}^{\eta}(\tau,\omega_{-}))^2}{R_{0}^{\eta}(\tau,\omega_{-})}\Big)d\omega_{-} .
\end{equation}
Substituting Eq.(\ref{R123-SR}) into Eq.(\ref{FI-SR}), we finally arrive at
\begin{eqnarray}
\label{FI-SR-0}
F_{\tau}^{\eta}(\tau)=\int\frac{\sqrt{2}(1-\gamma^2)V^2e^{-\omega_{-}^2/8\sigma_{-}^2}\omega_{-}^2\sin(\omega_{-}\tau)^2}{4(\pi \sigma_{-}^2)^{\frac{1}{2}}\Big(e^{\omega_{-}^2\eta_{\epsilon}^2/2}-V\cos(\omega_{-}\tau)\Big)\Big(e^{\omega_{-}^2\eta_{\epsilon}^2/2}\frac{1+3\gamma}{1-\gamma}+V\cos(\omega_{-}\tau)\Big)}d\omega_{-}.
\end{eqnarray}
Eq.(\ref{FI-SR-0}) can be used to analyze the sensitivity of the FI to the phase noise $\eta_{\epsilon}$ for spectrally resolved HOM interferometry. Under the ideal conditions of zero loss and perfect visibility, i.e., $\gamma=0,V=1$, we have
\begin{eqnarray}
\label{FI-SR-11}
F_{\tau,ideal}^{\eta}&&=\int\frac{\sqrt{2}e^{-\omega_{-}^2/8\sigma_{-}^2}\omega_{-}^2\sin(\omega_{-}\tau)^2}{4(\pi \sigma_{-}^2)^{\frac{1}{2}}(e^{\omega_{-}^2\eta_{\epsilon}^2}-\cos(\omega_{-}\tau)^2)}d\omega_{-}.
\end{eqnarray}
If there is no noise, i.e., $\eta_{\epsilon}=0$, then
\begin{eqnarray}
\label{FI-SR-11-1}
F_{\tau,ideal}^{\eta}&&=\int\frac{\sqrt{2}e^{-\omega_{-}^2/8\sigma_{-}^2}\omega_{-}^2}{4(\pi \sigma_{-}^2)^{\frac{1}{2}}}d\omega_{-} =4\sigma_{-}^2.
\end{eqnarray}

From Eq.(\ref{FI-SR-11-1}), it can be seen that we recover the QCRB under the ideal conditions of zero loss, perfect visibility, and zero noise, thus obtaining the sensing limit of the time delay $\tau$ based on spectrally resolved HOM interferometry. Moreover, it is worth noting that the FI with spectrally resolved detection does not depend on the time delay $\tau$ in this case. This means that one can achieve the QCRB at arbitrary time delay points, even if the time delay exceeds the coherence time of biphotons, referred to as the ambiguity-free dynamic range in \cite{CHENPRAPP2023}. However, this is not the case with spectrally non-resolved detection, where the FI always depends on the time delay even under the ideal conditions (see Eq.(\ref{FI-ideal-1})), and the QCRB can be realized only at the zero delay point.

Fig.\ref{Fig1} gives an example to display the sensitivity of quantum sensing based on HOM interferometry on the noise. In Fig.\ref{Fig1}, Fisher information, in units of $\omega_{p}^2$, as a function of $\omega_{p} \tau$ is plotted at different noise levels $\eta_{\epsilon}\omega_p$=0 (blue), 1 (orange), and 3 (green) for a fixed photon loss $\gamma$ and visibility $V$. $\omega_{p}$ is the center frequency of pump light, and we assume that $\omega_{p}=100\sigma_-$ in our calculations. It can be seen from Fig.\ref{Fig1}(a), under the ideal conditions of zero loss and perfect visibility, the FI is slightly sensitive to the phase noise, especially within the coherence time of biphotons for spectrally non-resolved (solid lines) detections and outside the coherence time of biphotons for the spectrally resolved (dashed lines) ones. For the FI with spectrally non-resolved (solid lines) detections in Fig.\ref{Fig1}(a), there appears an abrupt flip from the maxima to the minimum at the zero delay point when the phase noise changes from zero to other values, manifesting a significant difference between the ideal and practical situations. Furthermore, the spectrally resolved result in Fig.\ref{Fig1}(a) when $\eta_{\epsilon}=0$ (the blue dashed lines) saturates the QCRB at arbitrary delay points, while it holds only at the zero delay point for the spectrally non-resolved case (the blue solid lines), which is consistent with the results discussed from Eqs.(\ref{FI-ideal-1}) and (\ref{FI-SR-11-1}).

On the other hand, there are photon loss and imperfect visibility in practice, so the FI never saturates the QCRB at arbitrary delay points, whatever spectrally non-resolved or resolved detections are used. In other words, all the values of the FI are below the QCRB (bold black lines) in a practical scenario, as shown in Figs.\ref{Fig1}(b) and (c). From Fig.\ref{Fig1}(a) to Fig.\ref{Fig1}(c), we can see the FI gets smaller as the increase of the photon loss and the degeneration of the visibility. However, the FI is insensitive to the phase noise when $\gamma\neq0$ and $V\neq1$ (Figs.\ref{Fig1}(b) and (c)) both in the manners of spectrally resolved (dashed lines) and non-resolved (solid lines) detections. Physically, this insensitivity arises from the dependence of the coincidence probability on the biphoton frequency difference (low-frequency dependence) \cite{JIN-Optica-2018,Li-PRA-2023,OLT-2025}, enabling quantum sensing based on HOM interferometry to work in a high phase-noise environment. Furthermore, it is noted that the spectrally resolved results (dashed lines) are superior to the spectrally non-resolved ones (solid lines) in all cases, especially for the delay time that exceeds the coherence time of biphotons, indicating the spectrally resolved detection exhibits sensing advantages over the spectrally non-resolved one, as those in \cite{SR-PR2020,CHENPRAPP2023,OE-2025,Chen-OE-2025,SR-PRA2015,SR-OE2015,PRAPPLIED2020,JIN-2024,jin2016,OLT2023}.

It can be seen from Figs.\ref{Fig1}(b) and (c) that the maxima of the FI are shifted from the zero delay point to others under non-ideal conditions, but it is still possible to determine the optimal delay points, for example, by the maximum likelihood estimation (MLE), a widely used analytical technique for an estimator of $\tau$ \cite{SA2018,npj2019,CHENPRAPP2023}.

\section{\label{sec:3} Quantum sensing based on two-photon N00N state interferometry}
\subsection{\label{A'} Quantum sensing based on spectrally non-resolved N00N state interferometry}

Now, let us consider the sensitivity of the phase noise to quantum sensing based on two-photon N00N state interferometry under a noisy environment in an analogous manner as those in Part II.

The probability distributions for such an interferometry can be obtained \cite{JIN-Optica-2018,Li-PRA-2023,OLT-2025},
\begin{eqnarray}
\label{R123N00N}
&&R'_2(\tau)=\frac{1}{2}(1-\gamma)^2\Big(1+Ve^{-2\sigma_{+}^2(\tau-\epsilon)^2}\cos[\omega_p(\tau-\epsilon)+2\theta]\Big),
\nonumber\\
&&R'_1(\tau)=\frac{1}{2}(1-\gamma)^2\Big(\frac{1+3\gamma}{1-\gamma}-Ve^{-2\sigma_{+}^2(\tau-\epsilon)^2}\cos[\omega_p(\tau-\epsilon)+2\theta]\Big),\nonumber\\
&&R'_0(\tau)=\gamma^2.
\end{eqnarray}
where subscripts 0, 1, and 2 have similar definitions, as those in Part II. We use the superscript notation $'$ to represent the probability distributions of the N00N state interferometry. $\sigma_{+}$ denotes the RMS (root mean square) bandwidth of a Gaussian entangled state used, which is determined by the pump envelope function associated with biphoton frequency sum $\omega_+=\omega_s+\omega_i$ \cite{JIN-Optica-2018,Li-PRA-2023,OLT-2025}, and $\omega_p$ is the center frequency of the pump light. It is found that N00N state interferometry is influenced both by $\epsilon$ and $\theta$, which is different from the case of HOM interference only associated with $\epsilon$. After using the noise model in Eqs.(\ref{J}) and (\ref{RJ}), we have
\begin{eqnarray}
\label{R123N00N-noise}
&&R_2^{'\eta}(\tau)=\frac{1}{2}(1-\gamma)^2\Big(1+(V/\sqrt{A'})\cos(\omega_p\tau)e^{-2\eta_{\theta}^2-\frac{2\sigma_{+}^2\tau^2+\eta_{\epsilon}^2\omega_{p}^2/2}{A'}}\Big),
\nonumber\\
&&R_1^{'\eta}(\tau)=\frac{1}{2}(1-\gamma)^2\Big(\frac{1+3\gamma}{1-\gamma}-(V/\sqrt{A'})\cos(\omega_p\tau)e^{-2\eta_{\theta}^2-\frac{2\sigma_{+}^2\tau^2+\eta_{\epsilon}^2\omega_{p}^2/2}{A'}}\Big),\nonumber\\
&&R_0^{'\eta}(\tau)=\gamma^2.
\end{eqnarray}
where $A'=1+4\sigma_{+}^2\eta_{\epsilon}^2.$ If there is no noise, that is, $\eta_{\epsilon}=\eta_{\theta}=0$, then $A'=1$. Eq.(\ref{R123N00N-noise}) will reduce to the original formula \cite{JIN-Optica-2018,OLT-2025}.

The corresponding FI can be calculated as \cite{CHENPRAPP2023,PRAPP2023},
\begin{equation}
\label{FI-N00N}
F_{\tau}^{'\eta}=\sum_{i}\frac{(\partial_{\tau} R_i^{'\eta}(\tau))^2}{R_i^{'\eta}(\tau)}=\frac{(\partial_{\tau}R_2^{'\eta}(\tau))^2}{R_2^{'\eta}(\tau)}+\frac{(\partial_{\tau}R_1^{'\eta}(\tau))^2}{R_1^{'\eta}(\tau)}+\frac{(\partial_{\tau}R_0^{'\eta}(\tau))^2}{R_0^{'\eta}(\tau)}.
\end{equation}
 Substituting Eq.(\ref{R123N00N-noise}) into Eq.(\ref{FI-N00N}), we finally arrive at
\begin{eqnarray}
\label{FI-N00N-}
F_{\tau}^{'\eta}=\frac{V^2 (1-\gamma^{2})(4\sigma_{+}^2\tau \cos(\omega_p \tau)+\sin(\omega_p \tau)A'\omega_p)^2}{A'^{2}\Big(\frac{4V \gamma \cos(\omega_p \tau)A'^{\frac{1}{2}}}{1-\gamma} e^{\frac{2\sigma_{+}^2\tau^2+2A'\eta_{\theta}^2+\eta_{\epsilon}^2\omega_p^2/2}{A'}}+\frac{(1+3\gamma)A'}{1-\gamma}e^{\frac{4\sigma_{+}^2\tau^2+4A'\eta_{\theta}^2+\eta_{\epsilon}^2\omega_p^2}{A'}}-V^2 \cos(\omega_p \tau)^2\Big)} ,
\end{eqnarray}
Eq.(\ref{FI-N00N-}) can be used to analyze the sensitivity of the FI on the phase noise $\eta_{\epsilon}$ and $\eta_{\theta}$ for spectrally non-resolved N00N state interferometry. Under the ideal case of zero loss and perfect visibility, i.e., $\gamma=0,V=1$, Eq.(\ref{FI-N00N-}) can be simplified to
\begin{eqnarray}
\label{FI-ideal-}
F_{\tau}^{'\eta}=\frac{(4\sigma_{+}^2\tau \cos(\omega_p \tau)+\sin(\omega_p \tau)A'\omega_p)^2}{A'^{2}\Big(e^{(4\sigma_{+}^2\tau^2+4A'\eta_{\theta}^2+\eta_{\epsilon}^2\omega_p^2)/A'}- \cos(\omega_p \tau)^2\Big)},
\end{eqnarray}
If there is no noise, i.e., $\eta_{\epsilon}=\eta_{\theta}=0$, then
\begin{eqnarray}
\label{FI-ideal-2}
F_{\tau,ideal}^{'\eta}=\frac{(4\sigma_{+}^2\tau \cos(\omega_p\tau)+\sin(\omega_p\tau)\omega_p)^2}{e^{4\sigma_{+}^2\tau^2}-\cos(\omega_p\tau)^2},
\end{eqnarray}

This is the CRB for the N00N state interferometry. By evaluating the FI for this set of probabilities, we find that its upper bound is achieved under the ideal case at the zero delay point, given by
\begin{equation}
\label{FI0}
\lim_{\tau \to 0}F_{\tau,ideal}^{'\eta}=4\sigma_{+}^2+\omega_{p}^2.
\end{equation}
It is clear that we recover the QCRB under the ideal case at the zero delay point, thus obtaining the sensing limit of the time delay $\tau$ based on spectrally non-resolved N00N state interferometry.

\subsection{\label{B'} Quantum sensing based on spectrally resolved N00N state interferometry}

If the spectral resolution of the detector is high enough so that it can resolve the frequency of each photon, then, in this case, the coincidence probability can be obtained directly by removing the frequency integral of the original formula \cite{JIN-Optica-2018,Li-PRA-2023,OLT-2025},
\begin{eqnarray}
\label{R123-SR-}
&&R_2^{'\eta}(\tau,\omega_+)=\frac{(1-\gamma)^2}{4(2 \pi \sigma_{+}^2)^{\frac{1}{2}}}e^{-\frac{(\omega_+ -\omega_p)^2}{8\sigma_{+}^2}}\Big(1+V\cos[\omega_+(\tau-\epsilon)+2\theta]\Big),
\nonumber\\
&&R_1^{'\eta}(\tau,\omega_+)=\frac{(1-\gamma)^2}{4(2 \pi \sigma_{+}^2)^{\frac{1}{2}}}e^{-\frac{(\omega_+ -\omega_p)^2}{8\sigma_{+}^2}}\Big(\frac{1+3\gamma}{1-\gamma}-V\cos[\omega_+(\tau-\epsilon)+2\theta]\Big),\nonumber\\
&&R_0^{'\eta}(\tau,\omega_+)=\frac{\gamma^2}{2(2 \pi \sigma_{+}^2)^{\frac{1}{2}}}e^{-\frac{(\omega_+ -\omega_p)^2}{8\sigma_{+}^2}}.
\end{eqnarray}
After using the noise model in Eqs.(\ref{J}), and (\ref{RJ}), we have
\begin{eqnarray}
\label{R123-SR-noise}
&&R_2^{'\eta}(\tau,\omega_+)=\frac{(1-\gamma)^2}{4(2 \pi \sigma_{+}^2)^{\frac{1}{2}}}e^{-\frac{(\omega_+ -\omega_p)^2}{8\sigma_{+}^2}}\Big(1+Ve^{-2\eta_{\theta}^2-\eta_{\epsilon}^2\omega_{+}^2/2}\cos(\omega_+\tau)\Big),
\nonumber\\
&&R_1^{'\eta}(\tau,\omega_+)=\frac{(1-\gamma)^2}{4(2 \pi \sigma_{+}^2)^{\frac{1}{2}}}e^{-\frac{(\omega_+ -\omega_p)^2}{8\sigma_{+}^2}}\Big(\frac{1+3\gamma}{1-\gamma}-Ve^{-2\eta_{\theta}^2-\eta_{\epsilon}^2\omega_{+}^2/2}\cos(\omega_+\tau)\Big),\nonumber\\
&&R_0^{'\eta}(\tau,\omega_+)=\frac{\gamma^2}{2(2 \pi \sigma_{+}^2)^{\frac{1}{2}}}e^{-\frac{(\omega_+ -\omega_p)^2}{8\sigma_{+}^2}}.
\end{eqnarray}

\begin{figure}[th]
\begin{picture}(380,250)
\put(0,0){\makebox(375,240){
\scalebox{0.6}[0.6]{
\includegraphics{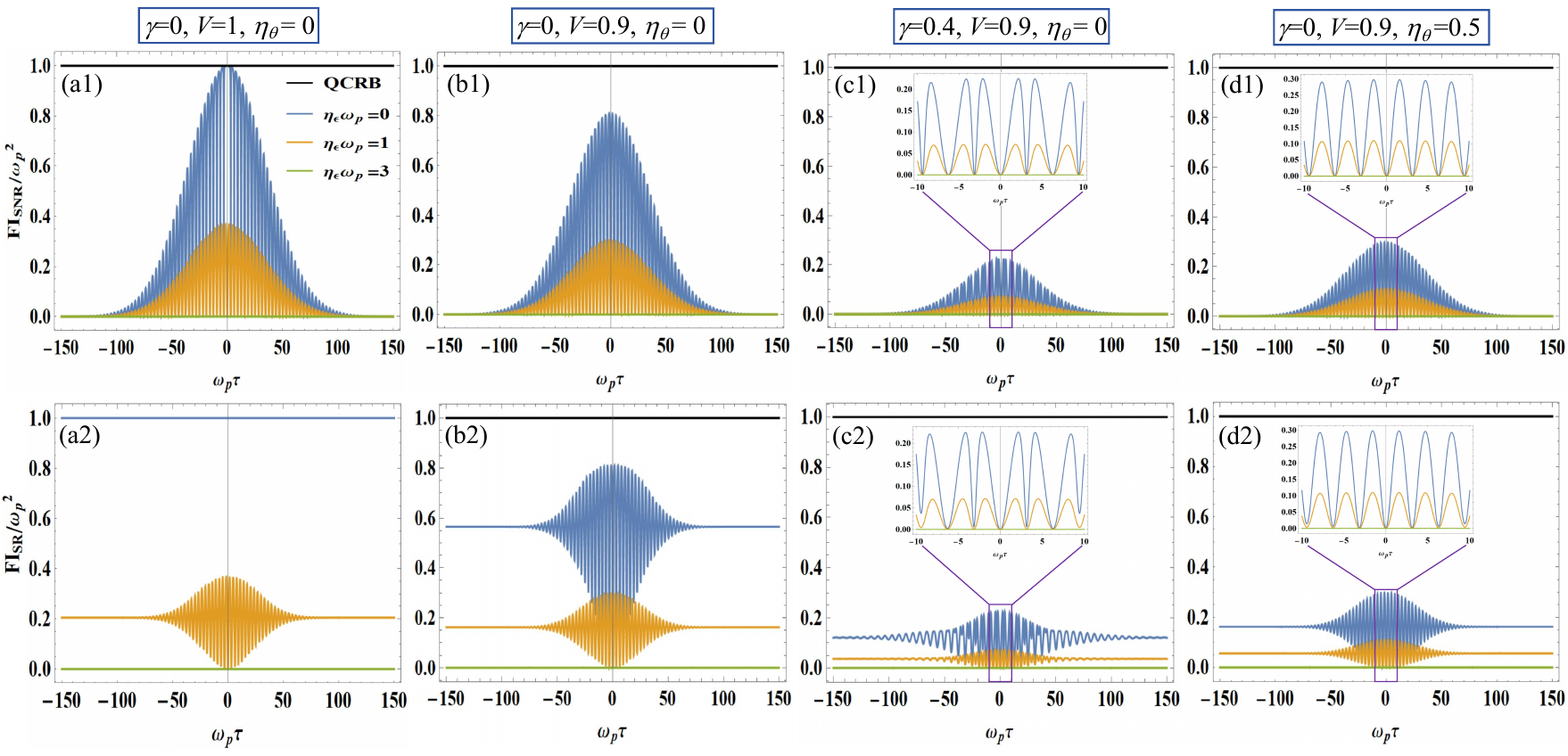}
}}}
\end{picture}
\caption{\label{Fig2}
Fisher information, in units of $\omega_{p}^2$, as a function of $\omega_{p} \tau$ for $\eta_{\epsilon}\omega_p$=0 (blue), 1 (orange), and 3 (green). The first and second rows represent spectrally non-resolved and resolved results, respectively. The insets provide enlarged views to show more details. The bold black lines represent the QCRB.}
\end{figure}

The corresponding FI can be calculated as \cite{CHENPRAPP2023,PRAPP2023}
\begin{equation}
\label{FI-SR-}
F_{\tau}^{'\eta}(\tau,\omega_+)=\int \Big(\frac{(\partial_{\tau}R_{2}^{'\eta}(\tau,\omega_+))^2}{R_{2}^{'\eta}(\tau,\omega_+)}+\frac{(\partial_{\tau}R_{1}^{'\eta}(\tau,\omega_+))^2}{R_{1}^{'\eta}(\tau,\omega_+)}+\frac{(\partial_{\tau}R_{0}^{'\eta}(\tau,\omega_+))^2}{R_{0}^{'\eta}(\tau,\omega_+)}\Big)d\omega_+ .
\end{equation}
Substituting Eq.(\ref{R123-SR-noise}) into Eq.(\ref{FI-SR-}), we finally arrive at
\begin{eqnarray}
\label{FI-SR-1}
F_{\tau}^{'\eta}=\int\frac{e^{-\frac{(\omega_+ -\omega_p)^2}{8\sigma_{+}^2}}V^2(1-\gamma^2)\sin(\omega_+\tau)^2\omega_{+}^2}{2(2\pi \sigma_{+}^2)^{\frac{1}{2}}\Big(e^{\omega_{+}^2\eta_{\epsilon}^2+4\eta_{\theta}^2}+V\cos(\omega_{+}\tau)\Big)\Big(e^{\omega_{+}^2\eta_{\epsilon}^2+4\eta_{\theta}^2}\frac{1+3\gamma}{1-\gamma}-V\cos(\omega_{+}\tau)\Big)}d\omega_+.
\end{eqnarray}
Eq.(\ref{FI-SR-1}) can be used to analyze the sensitivity of the FI to the phase noise $\eta_{\epsilon}$ and $\eta_{\theta}$ for spectrally resolved N00N state interferometry.

Under the ideal case of zero loss and perfect visibility, i.e., $\gamma=0,V=1$, we have
\begin{eqnarray}
\label{FI-SR-11-}
F_{\tau}^{'\eta}=\int\frac{e^{-\frac{(\omega_+ -\omega_p)^2}{8\sigma_{+}^2}}\sin(\omega_+\tau)^2\omega_{+}^2}{2(2\pi \sigma_{+}^2)^{\frac{1}{2}}\Big(e^{2(\omega_{+}^2\eta_{\epsilon}^2+4\eta_{\theta}^2)}-\cos(\omega_{+}\tau)^2\Big)}d\omega_+.
\end{eqnarray}
If there is no noise, i.e., $\eta_{\epsilon}=\eta_{\theta}=0$, then
\begin{eqnarray}
\label{FI-SR-11-2}
F_{\tau,ideal}^{'\eta}&&=\int\frac{e^{-\frac{(\omega_+ -\omega_p)^2}{8\sigma_{+}^2}}\omega_{+}^2}{2(2\pi \sigma_{+}^2)^{\frac{1}{2}}}d\omega_+ =4\sigma_{+}^2+\omega_{p}^2.
\end{eqnarray}
From Eq.(\ref{FI-SR-11-2}), it can be seen that we recover the QCRB under the ideal conditions of zero loss, perfect visibility, and zero noise, thus obtaining the sensing limit of the time delay $\tau$ based on spectrally resolved N00N state interferometry. Also, a similar case appears as HOM interferometry in Part II; that is, under the ideal conditions, the FI with spectrally resolved detection does not depend on the time delay $\tau$. This means that one can achieve the QCRB at arbitrary time delay points, even if the time delay exceeds the coherence time of biphotons. However, the FI always depends on the time delay even under the ideal conditions (see Eq.(\ref{FI-ideal-2})) for spectrally non-resolved detection, where the QCRB can be realized only at zero delay. This represents an advantage of a spectrally resolved scheme compared to a spectrally non-resolved one in N00N state interferometry.

As an example, Fig.\ref{Fig2} displays the sensitivity of quantum sensing based on N00N state interferometry on the noise. In Fig.\ref{Fig2}, Fisher information, in units of $\omega_{p}^2$, as a function of $\omega_{p} \tau$ is plotted at different noise levels $\eta_{\epsilon}\omega_p$=0 (blue), 1 (orange), and 3 (green) for a fixed photon loss $\gamma$ and visibility $V$. We assume that $\omega_{p}=100\sigma_+$ in our calculations. For the FI with spectrally non-resolved detections in Fig.\ref{Fig2}(a1), there also appears an abrupt flip from the maxima to the minimum at the zero delay point when the phase noise changes from zero to other values.Furthermore, the spectrally resolved result with $\eta_{\epsilon}=0$ (blue line) in Fig.\ref{Fig2}(a2) saturates the QCRB at arbitrary delay points, while it holds only at the zero delay point for the spectrally non-resolved case (Fig.\ref{Fig2}(a1)), which is consistent with the results discussed from Eqs.(\ref{FI-ideal-2}) and (\ref{FI-SR-11-2}). Likewise, it can be seen from the first to the third column in Fig.\ref{Fig2} that the FI gets smaller as the increase of the photon loss and the degeneration of the visibility as well, regardless of whether spectrally non-resolved or resolved detections are used. The fourth column in Fig.\ref{Fig2}(d1-d2) shows the results when a frequency-independent phase noise $\eta_{\theta}$ is introduced. Compared with the second column in Fig.\ref{Fig2}(b1-b2), the introduction of $\eta_{\theta}$ results in smaller FI, indicating the N00N state interferometry is sensitive to the phase noise $\eta_{\theta}$. It is worth noting that the period of all oscillations in Fig.\ref{Fig2} is basically the same, as seen clearly in the insets of Fig.\ref{Fig2} .

On the other hand, the spectrally resolved results (the second row) are always superior to the spectrally non-resolved ones (the first row) in all cases, especially for the delay time that exceeds the coherence time of biphotons, indicating the spectrally resolved detection exhibits advantages over the spectrally non-resolved one in quantum sensing based on N00N state interferometry under the noise environment. The conclusions above are the same as those obtained with HOM interferometry in Part II (Section B).

However, it can be seen from Fig.\ref{Fig2} that the FI is extremely sensitive to the phase noise in all cases regardless of whether the conditions are ideal or not and regardless of whether spectrally non-resolved or resolved detections are used. Physically, this sensitivity arises from the dependence of the coincidence probability on the biphoton frequency sum (high-frequency dependence) \cite{JIN-Optica-2018,Li-PRA-2023,OLT-2025}, enabling quantum sensing based on N00N state interferometry to work in a low phase-noise environment. 

\section{\label{sec:conclude}discussions and CONCLUSIONS}
By comparing Fig.~\ref{Fig1} and Fig.~\ref{Fig2}, we can observe that the peak height of the FI in Fig.~\ref{Fig1} is four orders of magnitude smaller than that of the ones in Fig.~\ref{Fig2}, reflecting the extreme-low phase sensitivity of the HOM interferometry. As $\eta_\epsilon$ increases, the FI remains nearly unchanged, indicating that the HOM interferometry is insensitive to the noise in practice. This suggests that the sensing application of HOM interferometry is more appropriate for high-noise scenarios. In addition to the insensitivity to the noise, the HOM interferometry is also robust against the turbulent atmosphere \cite{PRA-2018}, making it a versatile candidate for the practical application of various quantum technologies. In contrast, the FI in Fig.~\ref{Fig2} exhibits an extremely high peak around $\tau = 0$, reflecting the high phase sensitivity of the N00N state interferometry. However, as $\eta_\epsilon$ increases, the peak amplitude rapidly decays to zero, indicating that the N00N state interferometry is extremely sensitive to the noise. This suggests that the sensing application of N00N state interferometry is more appropriate for low-noise scenarios. Physically, this insensitivity (HOM interferometry) or sensitivity (N00N state interferometry) to the phase noise arises from the dependence of the coincidence probabilities on the biphoton frequency difference (low-frequency dependence) or sum (high-frequency dependence). Moreover, the spectrally resolved scheme is a better strategy for the two interferometry because it offers higher Fisher information, shorter measurement times, and an ambiguity-free dynamic range, especially for the scope that exceeds the coherence time of biphotons.

The time delay is merely one example of parameters in numerous quantum sensing applications. It can also be transformed conveniently into other sensing parameters, such as tiny phase, displacement, angle, and refractive index, where the sensitivity analysis of the two interferometry to the noise holds as well. The analysis of the sensitivity of HOM and N00N state interferometry to the phase noise can also be extended to other two-photon interference-based sensing that depends on two-photon frequency difference or sum. Of course, it is possible to incorporate both the phase-sensitive and phase-insensitive interference into a single interferometer \cite{Li-PRA-2023,OLT-2025,Li-PRA-2024,Li-APLP-2025} for versatile sensing applications, leaving it as our future work. 

In conclusion, we have studied comprehensively the sensitivity of quantum sensing based on HOM and N00N state interferometry on the phase noise. It is found that HOM (N00N state) interferometry in practice is insensitive (sensitive) to the phase noise, suggesting that its sensing application is appropriate for high-noise (low-noise) scenarios. This conclusion holds both in the manners of spectrally non-resolved and resolved detections. Moreover, we have obtained the ultimate sensing limit of these two interferometries under ideal conditions of zero loss and perfect visibility. However, this limit can be saturated at arbitrary delay points for spectrally resolved detection but only at the zero delay point for the spectrally non-resolved one. Also, the spectrally resolved scheme outperforms the spectrally non-resolved one, which opens up a new way for future high-precision quantum sensing. The findings provide an optimal strategy for the practical applications of two-photon interferometric sensing in diverse noise environments.

Note: The detailed derivation of the coincidence probability for HOM and N00N state interferometry without noise can be found in \cite{JIN-Optica-2018,OLT-2025}.

\begin{acknowledgments}
This work has been supported by National Natural Science Foundation of China (12074309, 92365106, 12074299), Scientific Research Program Funded by Shaanxi Provincial Education Department (Program No.24JR034), and the Youth Innovation Team of Shaanxi Universities.
\end{acknowledgments}


\end{CJK*}
\end{document}